\documentclass[aps,prb,twocolumn,showpacs]{revtex4-1}
\usepackage{graphicx}
\usepackage{latexsym}
\usepackage{amsmath}
\usepackage{txfonts}
\usepackage{helvet}

\begin{document}

\title{Spin disorder in an Ising honeycomb chain cobaltate} 

\author{Hirokazu Igarashi}
\author{Yasuhiro Shimizu}
\email{yasuhiro@iar.nagoya-u.ac.jp}
\author{Yoshiaki Kobayashi}
\author{Masayuki Itoh}
\affiliation{Department of Physics, Graduate School of Science, Nagoya University, Furo-cho, Chikusa-ku, Nagoya 464-8602, Japan.}
\date{\today}

\begin{abstract}
	We report a new member of the spin-disordered honeycomb lattice antiferromagnet in a quasi-one-dimensional cobaltate Ba$_3$Co$_2$O$_6$(CO$_3$)$_{0.7}$. 
	Resistivity exhibits as semimetallic along the face-sharing CoO$_6$ chains. 
	Magnetic susceptibility shows strongly anisotropic Ising-spin character with the easy axis along the chain due to significant spin-orbit coupling and a trigonal crystal field. 
	Nevertheless, $^{135}$Ba NMR detects no indication of the long-range magnetic order down to 0.48 K. 
	Marginally itinerant electrons possess large entropy and low-lying excitations with a Wilson ratio $R_W$ = 116, which highlight interplays of charge, spin, and orbital in the disordered ground state. 
\end{abstract}

\pacs{72.15.Jf, 75.40.-s, 75.10.Kt, 75.30.Gw}

\keywords{}

\maketitle

%\section{Introduction}
%1st paragraph
	%
	Quantum disorder states have attracted material scientists since proposals of residual entropy\cite{Wannier} and resonating valence bond\cite{Anderson} on geometrically frustrated spin lattices. 
	In the absence of frustration, a disorder state is realized due to low-dimensional fluctuations. 
	In dimensions larger than two, a honeycomb lattice has the smallest coordination number and hence largest fluctuations. 
	Realization of the quantum spin liquid on the honeycomb lattice has attracted great theoretical interests.\cite{Meng, Kitaev}
	In a strong coupling limit, however, Heisenberg models with nearest neighbor interactions give N${\rm \acute{e}}$el order ground states.\cite{Wannier, Diep, Oitmaa, Andrews} 
	Consideration of second and third neighbor exchange interactions induces emergent quantum liquids.\cite{Moessner, Lee, Hermele, Meng, Okumura, Clark} 
	Low-lying excitations from the ground state are discussed with respect to gapped\cite{Meng} or gapless\cite{Lee, Hermele} spinon.
	In the presence of moderate electron correlations and spin-orbit coupling,  the Mott transition occurs from a (topological) spin liquid to Dirac semimetal or a topological insulator.\cite{Kitaev, Shitade, Chaloupka, Pesin} 

	Experimental examples of the honeycomb lattice have been recently found in some transition-metal oxides such as Bi$_3$Mn$_4$O$_{12}$(NO$_3$),\cite{Smirnova} Ba$_3$CuSb$_2$O$_9$,\cite{Nakatsuji} and Na$_2$IrO$_3$.\cite{Singh} 
	However, the ground state undergoes spin glass for the manganite\cite{Onishi} and a gapped spin-liquid for the cuprate,\cite{Quilliam} and a N${\rm\acute{e}}$el order in the iridate. 
	Here we add a new example of the honeycomb lattice antiferromagnet, a quasi-one-dimensional (quasi-1D) cobaltate Ba$_3$Co$_2$O$_6$(CO$_3$)$_{0.7}$ (Ba326) with gapless low-lying excitations. 
	In contrast to the previous examples, the honeycomb lattice is constructed by itinerant Ising chains for Ba326. 
	The crystal structure belongs to hexagonal $P\bar{6}$ and consists of CoO$_6$ and CO$_3$ chains running along the $c$ axis,\cite{Boulahya, Iwasaki} as shown in Fig. 1(a). 
	The magnetic CoO$_6$ chains form the honeycomb lattice in the $ab$ plane, and the nonmagnetic CO$_3^{2-}$ chains are located in a tunnel of the honeycomb [Fig. 1(b)]. 
	The concentration of CO$_3$ ions was determined from the oxygen contents by inert gas fusion-infrared absorption analysis.\cite{Iwasaki} 
	The incommensurate stoichiometry (possibly disordered) gives a partially-filled conduction band with the nominal valence of Co$^{3.7+}$ ($3d^{5.3}$). 
	The chain structure and transport properties of Ba326 are analogous to those of Ca$_3$Co$_2$O$_6$ (Ca326)\cite{Kageyama} and Ca$_3$CoRhO$_6$ (Ref.\onlinecite{Niitaka}) with partially-disordered states on triangular-lattice chains. 
	However, the low-temperature magnetic and electric properties have not been investigated for Ba326. 
	Iwasaki {\it et al.} reported the metallic conductivity $\sigma$ and high Seebeck coefficient $S_{\rm e}$ above 300 K.\cite{Iwasaki} 
	The figure of merit $ZT = 0.51 \times 10^{-5}$ K$^{-1}$ ($Z = \sigma S_{\rm e}^2 \kappa^{-1}$, $\kappa$: thermal conductivity) is comparable to those of Ca326 and Na$_{0.75}$CoO$_2$ at 300 K.\cite{Iwasaki} 

	\begin{figure}
	\includegraphics[scale=0.5]{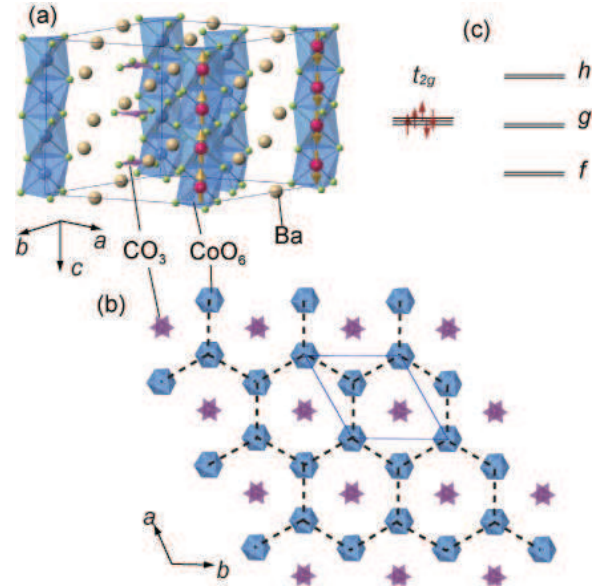}
	\caption{\label{Fig1} 
	(a) Crystal structure of Ba326 consisting of independent chains of CoO$_6$, closed-shell Ba, and CO$_3$ along the $c$ axis. 
	Non-stoichiometric CO$_3$ occupies 90\% in the crystal structure, as determined from the chemical analysis and the X-ray diffraction measurement.\cite{Iwasaki} 
	The arrows illustrate spins with ferromagnetic correlations along the chain and antiferromagnetic one in the $ab$ plane. 
	(b) The $c$-axis projected crystal structure where CoO$_6$ chains form a honeycomb lattice in the $ab$ plane and CO$_3$ chains are placed inside honeycomb framework. 
	(c) $t_{2g}$ orbital occupations of Co$^{4+}$ and the spin-orbital levels consisting of three sets of Kramers doublets ($f$, $g$, $h$) in presence of spin-orbit coupling and trigonal field, where a hole is occupied into the highest $h$ levels.\cite{Maekawa} 
	}
	\end{figure}
	
	In a trigonal crystal field of the face-sharing CoO$_6$,  a $t_{2g}$ triplet (an effective angular momentum $l$ = 1) splits into a lower lying $|l_z = 0 \rangle = a_{1g}$ singlet and a higher lying $|l_z = \pm 1 \rangle = e^\prime_g$ doublet. 
	The residual orbital degeneracy for Co$^{4+}$ ($d^5$) leads to spin-orbit coupling $\lambda {\bf s} \cdot {\bf l}$ with a coupling constant $\lambda$, comparable to the crystal field splitting ($\sim$ 0.1 eV) and exchange couplings.\cite{Maekawa} 
	Then the spin-orbital levels [Fig. 1(c)] are expressed by coherent mixtures of different orbital and spin states.\cite{Maekawa} 
	Under anisotropic electron correlations and frustration, the magnetism and transport properties carried by spin-orbital entangled states may have emergent features, as extensively investigated in $5d$ compounds providing a strong $\lambda$ limit.\cite{Shitade, Chaloupka} 
	In this respect, Ba326 is linked to a charge-spin-orbital coupled system on the honeycomb lattice. 

	In this paper, we investigate the ground state of Ba326 through transport, thermodynamic, and magnetic measurements. 
	The charge transport is confined into the $c$-axis chain and becomes weakly localized in the ground state. 
	The magnetic susceptibility exhibits Curie-Weiss-like temperature dependence with the anisotropic Weiss temperature implying anisotropic exchange interactions between Ising spins. 
	We found no indication of the long-range magnetic order and structural transitions in NMR and specific heat measurements. 
	The reasons are discussed with respect to spin-orbit interactions in the itinerant media. 

	Single crystals of Ba326 were prepared by a K$_2$CO$_3$-BaCl$_2$ flux method using a mixture of BaCO$_3$, Co$_3$O$_4$, K$_2$CO$_3$, and BaCl$_2$ reagents at 1273 K.\cite{Iwasaki} 
	The $c$ axis of needle-shape crystals was determined by X-ray diffraction patterns with peaks at $(00l)$ ($l$ : integer). 
	The typical size of the crystal was 1 mm $\times$ 1 mm $\times$ 3 mm. 
	To conduct resistivity and thermopower measurements, gold wires were attached to the sample after breaking an oxidized insulating surface. 
	The magnetic susceptibility, NMR, and specific heat were measured for aligned single crystals to gain sensitivity. 
	$^{135}$Ba NMR measurements were performed in constant magnetic fields $H_0$ = 9.4 T (1.9-200 K) and 8.5 T (0.48-2.5 K) parallel to the $c$ axis. 
	$^{135, 137}$Ba NQR spectrum was not detected in the frequency range of 6-50 MHz at zero field and hence the nuclear quadrupole frequency $\nu_Q$ was expected lower than 6 MHz. 
	The spin-echo intensity was not large enough to obtain the nuclear spin-lattice relaxation rate $1/T_1$ due to the small sample amount. 
	Instead, the nuclear spin-spin relaxation rate $1/T_2$ was obtained from the single-exponential decay of integrated spin-echo intensities as a function of pulse interval time $2\tau$. 

	\begin{figure}
	\includegraphics[scale=0.75]{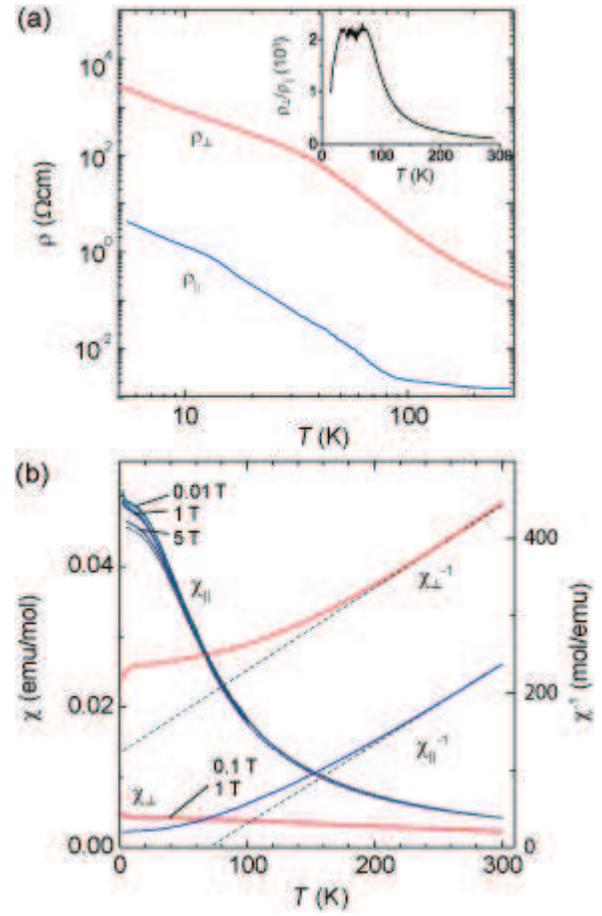}
	\caption{\label{Fig2} 
	(a) Resistivity measured parallel, $\rho_\parallel$, and perpendicular, $\rho_\perp$, to the $c$ axis for the identical crystal of Ba326. 
	Inset: the resistivity anisotropy $\rho_\perp$/$\rho_\parallel$ plotted against temperature. 
	(b) Magnetic susceptibility under magnetic fields (0.01, 1, 5 T) parallel, $\chi_\parallel$, or (0.1, 1 T) perpendicular, $\chi_\perp$, to the $c$ axis for the zero-field-cooling (ZFC, solid curves) and field cooling (FC, dotted curves) measurements (the left-hand axis). 
	$\chi^{-1}$ at 1 T fits to the Curie-Weiss law (dotted lines) in the $T$ range 220-300 K (the right-hand axis). 
	}
	\end{figure}

	Figure 2 shows the temperature $T$ dependence of resistivity and magnetic susceptibility below 300 K. 
	The parallel resistivity $\rho_\parallel$ measured along the $c$ axis is conducting with weak $T$ dependence, while the perpendicular one $\rho_\perp$ behaves semiconducting down to 5 K. 
	$\rho_\parallel$ seems to depend on the current path because of the insulating surface and basically metallic at high temperatures. 
	Resistivity anisotropy ($\rho_\perp /\rho_\parallel \sim 10^2$ at 290 K) increases on cooling $T$ ($\sim 2 \times 10^3$ at 50 K) [inset of Fig. 2(a)] and shows the quasi-1D electronic structure, as expected from the anisotropic crystal structure. 
	Below 100 K, $\rho_\parallel$ increases by three orders of magnitude down to 5 K. 
	Since $\rho_\perp$ exhibits no anomaly around 100 K, the $\rho_\parallel$ increase unlikely comes from a phase transition due to charge and orbital ordering. 
	Considering the incommensurate potential from CO$_3$ ions, weak localization is a possible source of the resistivity increase for the 1D system. 
	However, $\rho_\parallel (T)$ does not follow a normal activation type or a variable range hopping model but behaves close to a power law, $\sim T^{-2.5}$, below 80 K. 

	\begin{figure}
	\includegraphics[scale=0.45]{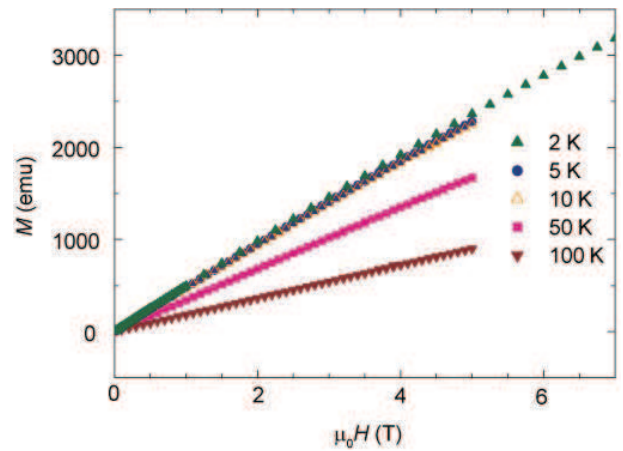}
	\caption{\label{Fig3} 
	Magnetization of Ba326 as a function of magnetic field parallel to the $c$ axis at various temperatures. 
	}
	\end{figure}
	
	The magnetic susceptibility in Fig. 2 (b) also displays highly anisotropic behavior. 
	The parallel susceptibility $\chi_{\parallel}$ increases with decreasing $T$ and becomes more than 10 times larger than the perpendicular one $\chi_{\perp}$ at 20 K. 
	It manifests the strong Ising nature due to spin-orbit coupling. 
	$\chi_{\parallel}^{-1}$ and $\chi_{\perp}^{-1}$ show linear $T$ dependence above 200 K with nearly the same slop. 
	Assuming the Curie-Weiss law $\chi_i^{-1} = (T - \Theta_i)/C_i^\prime$ ($i = \parallel$, $\perp$) above 200 K, we obtained the Curie constant $C^\prime_\parallel$ = 0.48 emu/(Co-mol K) and Weiss temperatures $\Theta_\parallel$ = 69 K for $\chi_{\parallel}^{-1}$, and $C^\prime_{\perp}$ = 0.46 emu/(Co-mol K) and $\Theta_\perp$ = $-100$ K for $\chi_{\perp}^{-1}$. 
	$\Theta$ should be isotropic in the Heisenberg model and expressed by $\Theta = (2J_\parallel + 3J_\perp)g_i^2/4k_{\rm B}$, where $J_\parallel$ and $J_\perp$ are exchange couplings parallel and perpendicular to the chain, respectively. 
	The anisotropy likely arises from the trigonal crystal field and spin-orbit coupling, as well as the exchange anisotropy. 
	The effective Curie constant has been evaluated for Co$^{2+}$ as $C^\prime_i = (N\mu_{\rm B}^2 /k_{\rm B})(g_i^2+B_i k_{\rm B}T/\lambda )$, where $N$ is the Avogadro's number, $g_i$ the thermally averaged g-factor, $\mu_{\rm B}$ the Bohr magneton, and $B_i$ the thermally averaged second-order Zeeman coefficient.\cite{Lines}
	Postulating the same model for Co$^{4+}$, the nearly isotropic $C^\prime_i$ despite the significant spin-orbit coupling is attributable to cancellation of the anisotropy in $g_i^2$ and $B_i k_{\rm B}T/\lambda $. 
	For the low-spin Co$^{4+}$ ($S = 1/2$) and Co$^{3+}$ ($S = 0$), the effective Curie constant  $C^\prime_i = (N\mu_{\rm B}^2 g^2 \sqrt{S(S+1)}/3k_{\rm B})$ gives the averaged g-value, $g = 2.7$. 

	The deviation of $\chi_{\parallel}^{-1}$ and $\chi_{\perp}^{-1}$ from the Curie-Weiss law below 150 K points to significant short-range correlation and spin-orbit coupling. 
	In the classical Ising spin system with ferromagnetic interactions, the long-range magnetic order occurs at $T_c \sim \Theta$ as three-dimensional coupling sets in. 
	Hence there are strong frustration and/or fluctuations against the ordering. 
	Further decreasing $T$, $\chi_{\parallel}$ and $\chi_{\perp}$ exhibit no appreciable difference in the zero-field-cooled and field-cooled data at low fields. 
	The result rules out a possible spin glass state. 
	As shown in Fig. 3, the magnetization $M$ is nearly linear to the external magnetic field parallel to the chain down to 2 K. 
	A weak field-dependence in $\chi_{\parallel}$  below 50 K implies that $M$ begins to saturate at high fields and low temperatures. 
	This robust paramagnetic behavior is surprising for the longitudinal magnetic field parallel to the Ising easy axis, because the analogous compound Ca326 with $S$ = 2 exhibits field-induced ferrimagnetic transitions.\cite{Kageyama, Shimizu} 

	\begin{figure}
	\includegraphics[scale=0.5]{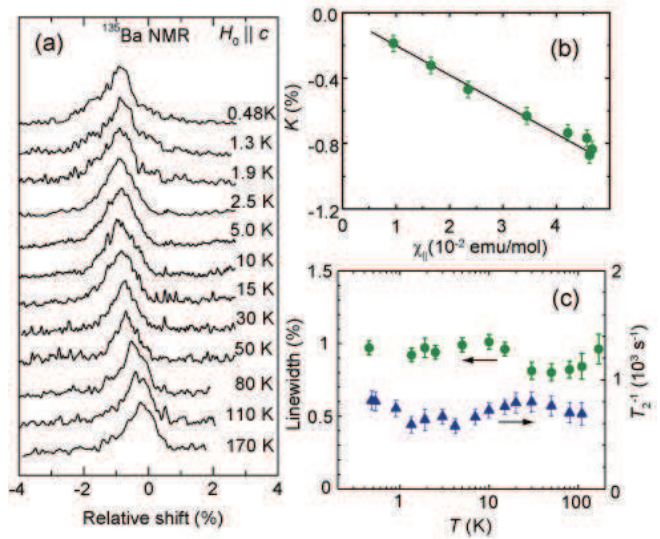}
	\caption{\label{Fig4} 
	(a) $^{135}$Ba NMR spectra for aligned single crystals of Ba326, measured at $H_0$ = 9.4 ($T >$ 2 K) and 8.5 ($T <$ 2 K) T parallel to the $c$ axis. 
	(b) Knight shift $K$ versus magnetic susceptibility $\chi_{\parallel}$ plot. 
	(c) $T$ dependence of the linewidth (left axis) defined by the full-width at half maximum intensity and the nuclear spin-spin relaxation rate $T_2^{-1}$ (right axis). 
	}
	\end{figure}

	NMR is utilized as a sensitive microscopic probe of magnetic ordering. 
	$^{135}$Ba NMR spectrum measures transfer hyperfine and dipole fields from Co moments as a resonance shift and broadening. 
	The crystal structure of Ba326 has three Ba sites.  
	In a magnetic field $H_0 \parallel c$, we observed a single broad spectrum. 
	The linewidth comes from the three Ba sites with the slightly different hyperfine coupling and electric field gradient under the disordered potential from CO$_3$. 
	The nuclear quadrupole satellites were not observed in the measured frequency range $\sim$ 3 MHz. 
	The spectrum exhibits a negative hyperfine shift $K$ on cooling [Fig. 4(a)]. 
	$K$ well scales to $\chi_{\parallel}$ with a hyperfine coupling constant $A$ = $-0.21$ T/$\mu_{\rm B}$ [Fig. 4(b)]. 
	The paramagnetic shift down to 2 K confirms the absence of spontaneous local fields. 
	A slight change in the line shape below 2 K comes from a misalignment of the external field by $\sim $ 5$^\circ$ on switching to a $^3$He cryostat, which does not influence the linewidth defined by full-width at half-maximum. 
	As seen in Fig. 4(c), the linewidth is nearly independent of $T$ and biased by the inhomogeneous distribution of the electric field gradient. 
	The weak temperature dependence of 0.2\% ($\sim$0.2 T) below 30 K should be a paramagnetic effect proportional to the Knight shift. 
	If it were due to spontaneous magnetic moments, it is no more than 0.1$\mu_{\rm B}$ using the obtained $A$. 
	The temperature dependence of $1/T_2$ is also displayed in Fig. 4(c), which is sensitive to spin fluctuations. 
	$1/T_2$ is nearly invariant against $T$ [Fig. 4 (c)] without the indication of long-range magnetic ordering. 
	The behavior is distinct from quantum antiferromagnts showing a divergent increase at low temperatures.\cite{Takigawa}

	\begin{figure}
	\includegraphics[scale=0.45]{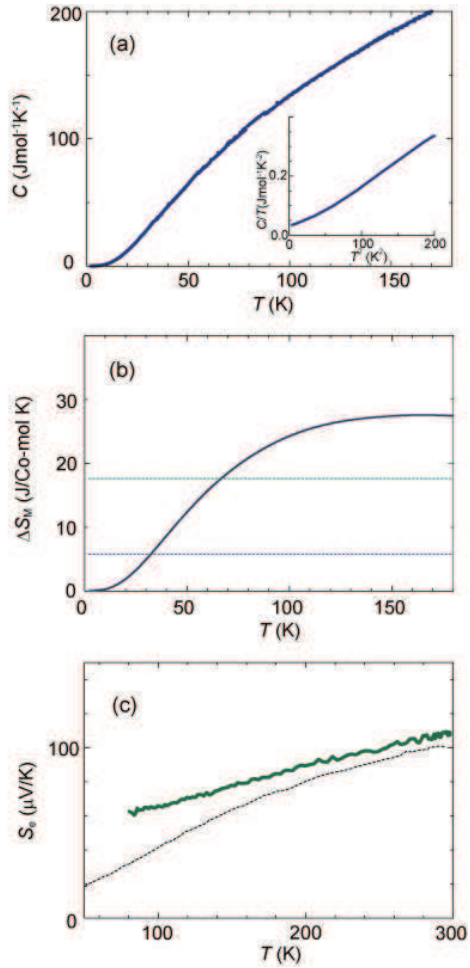}
	\caption{\label{Fig5} 
	(a) Specific heat $C$ of Ba326 measured at zero field. 
	Inset: $C/T$ plotted against $T^2$, where the extrapolation to $T = 0$ gives the $T$-linear coefficient $\gamma$. 
	(b) Temperature dependence of the magnetic entropy $\Delta S_{\rm M}$ obtained from the $T$-integral of $C/T$, where the lattice contribution to $C$ was subtracted by using $C$ of nonmagnetic Ca$_4$PtO$_6$. 
	The dotted lines denote the entropy expected for spin $S = 1/2$, $R$ln2 (blue), and for extra orbital degrees of freedom (green, see text for the detail). 
	(c) Seebeck coefficient $S_e$ measured parallel to the chain in Ba326. 
	A dotted curve shows the data of Na$_{0.75}$CoO$_2$ (Ref. \onlinecite{Terasaki} ) for comparison.
	}
	\end{figure}

	The specific heat $C$ displays no anomaly due to phase transitions down to 2 K, as displayed in Fig. 5(a). 
	The $T$-linear coefficient $\gamma$ is evaluated from the $C/T$ intercept plotted against $T^2$ [the inset of Fig. 5 (a)]. 
	The finite $\gamma$ indicates low-lying gapless excitations. 
	The anomalous $\gamma$ has been also observed in the spin-chain system such as Ca326 (Ref. \onlinecite{Hardy}) and Sr$_5$Rh$_4$O$_{12}$ (Ref. \onlinecite{Cao}) with magnetic ordering. 
	In Ba326 with itinerant and localized spins without ordering, $\gamma$ measures the charge and spinon density of states.\cite{Balents, Chen}
	The obtained $\gamma = 15$ mJ/(Co-mol K$^2$) is comparable to that of the thermoelectric cobaltate Na$_{0.75}$CoO$_2$, 16 mJ/(Co-mol K$^2$), with the enhanced density of states.\cite{Ong}  
	Considering the large residual $\chi$ without long-range ordering, local moments are expected to dominate low-energy excitations. 
	The Wilson ratio $R_W = (\pi^2/3)(\chi/\mu_{\rm B}^2)/(\gamma/k_{\rm B}^2)$ is evaluated as $R_{W \parallel} = 116$ and $R_{W \perp} =10$ using $\chi_{\parallel}$ = 0.047 and $\chi_{\perp} = 0.0043$ emu/mol at 20 K. 
	As known in some spin liquid candidates, $R_W$ is close to unity for organic systems and enhanced in presence of the spin-orbit coupling\cite{Balents} such as the hyperkagome lattice Na$_4$Ir$_3$O$_8$ (Ref. \onlinecite{Okamoto}) and the diamond lattice FeSc$_2$S$_4$ (Ref. \onlinecite{Fritsch}). 
	The enhanced $R_W$ highlights the significant role of spin-orbit coupling in the spinon excitations.\cite{Kim} 

	From the specific heat data, the magnetic entropy $\Delta S_{\rm M}$ was obtained after subtracting the lattice contribution [Fig. 5(b)]. 
	In a paramagnetic Mott insulator with $S$ = 1/2, $\Delta S_{\rm M}$ reaches $\Delta S_{1/2}$ = $R$ln($2S+1$) = $R$ln2 = 5.76 ($R$: the gas constant) with increasing $T$. 
	We obtained $\Delta S_{\rm M}$ far exceeding $\Delta S_{1/2}$. 
	The extra contribution is in part attributed to orbital degrees of freedom. 
	Here the spin-orbital degeneracy is at most 6 for Co$^{4+}$ ($d^5$, $S$ = 1/2) and 18 for Co$^{3+}$ ($d^6$, $S$ = 1),\cite{Koshibae} leading to $\Delta S_{\rm M}$ = $R$(0.7ln6+0.3ln18) = 17.6, as shown in the dotted line in Fig. 5(b). 
	Additional valence fluctuations should be also contribute to the excess entropy. 
	Another manifestation of the high entropy is the Seebeck coefficient $S_{\rm e}$ defined by thermopower per $T$, as shown in Fig. 5(c). 
	The positive $S_{\rm e}$ points to hole conductivity, as expected for the $3d^{5.3}$ occupation in the $t_{2g}$ orbitals. 
	Similar $T$ dependence was observed in a typical thermoelectric material Na$_{0.75}$CoO$_2$ (Ref. \onlinecite{Terasaki}). 
	Since $S_{\rm e}$ relates to entropy per a conduction electron, the large $S_{\rm e}$ is consistent with the specific heat. 

	Now we discuss the microscopic origin for the exotic magnetic and transport properties of Ba326. 
	The notable feature is disordering of local moments despite the strong Ising anisotropy. 
	Taking account of the local trigonal distortion along the chain, the $S \sim 1/2$ local moment comes from the orbital dependent localization. 
	Without spin-orbital coupling, the $t_{2g}$ multiplet consists of $a_{1g} = (yz+zx+xy)/{\surd 3}$ and $e_g^\prime = (e^{\pm i2\pi/3}yz + e^{\mp i2\pi/3}zx + xy)/{\surd 3}$ orbitals. 
	The $a_{1g}$ orbital elongated along the chain forms the direct $d$-$d$ overlap and hence a partially-filled 1D conduction band, whereas the orphan $e_g^\prime$ orbitals can carry local moments. 
	The double-exchange interaction between $e_g^\prime$ spins via the itinerant $a_{1g}$ orbital causes ferromagnetic correlations along the chain. 
	The direct $d$-$d$ exchange is expected to be much larger than the interchain superexchange. 
	Along this line, a promising ground state is the N${\rm \acute{e}}$el order for ferromagnetically aligned giant spins on the bipartite honeycomb lattice.\cite{Oitmaa, Andrews} 

	To explain the unexpected disorder state, we consider three possible origins. 
	First one is the spin-orbit coupling, as manifested in the Ising anisotropy. 
	It gives rise to the entanglement of the itinerant $a_{1g}$ and localized $e_g^\prime$ spins.\cite{Bhattacharjee} 
	Such a complex admixture may induce Kondo coupling and enhance quantum fluctuations.\cite{Chen} 
	Indeed, the magnetic susceptibility levels off as the resistivity goes up, analogous to the Kondo system. 
	The second reason is the presence of itinerant electrons that dislike the N${\rm \acute{e}}$el order state. 
	Although the system becomes weakly localized at low temperatures, the itinerancy of $a_{1g}$ spin is robust due to the strong ferromagnetic correlation as seen in the good conductivity. 
	The third one is geometrical frustration arising from the next neighbor interactions $J^\prime_\perp$ on the honeycomb lattice. 
	The critical value of $J^\prime_\perp$ is evaluated as only $J^\prime_\perp = 0.08J_\perp$ for inducing a spin liquid phase in a Heisenberg system.\cite{Clark, Okumura} 
	However, the present system has the marginally itinerant character with the Ising anisotropy, and no theoretical model has been reported so far.  
	Although the ground state of Ba326 is not fully understood yet, the quantum fluctuations likely arise from these complex interplays of spin, charge, and orbital. 
	To give further insight into the physics in Ba326, it will be important to control the ground state by chemical doping or pressure. 

	In conclusion, we have investigated the ground state properties of the thermoelectric cobaltate Ba$_3$Co$_2$O$_6$(CO$_3$)$_{0.7}$ with the honeycomb lattice. 
	The compound is found to be a rare example of the honeycomb-lattice Ising chain system with the effective spin $S = 1/2$. 
	The magnetic susceptibility, NMR, and specific heat measurements show no indication of long-range magnetic ordering down to low temperatures. 
	The low-lying excitations with the large Wilson ratio suggest the spin-orbit coupled itinerant system. 
	The result will open further theoretical studies of quantum liquid with moderate electron correlations and spin-orbit coupling. 

	The authors thank Inoue S. for technical assistance, and valuable discussion with M. Tsuchiizu, Y. Motome, N. P. Ong, A. Sondi, and J. Moore.  
	This work was financially supported by the Grants-in-Aid for Scientific Research (No.25610093, 23225005, and 24340080) from JSPS, and the Grant-in-Aid for Scientific Research (No. 22014006) on Priority Areas "Novel State of Matter Induced by Frustration" from the MEXT. 

\bibliography{Ba326}

\end{document}